\documentclass[10pt]{article}
\usepackage[dvips]{graphicx}

\usepackage{epsfig}
\usepackage{graphicx}
\usepackage{indentfirst}
\usepackage{amsmath}
\usepackage{amsfonts}
\usepackage{amssymb}

\begin{document}

\begin{center}
{\Large\bf Realistic $f(T)$ model describing the de Sitter epoch of the dark energy dominated universe  }\\

\medskip

S. B. Nassur$^{(a)}$\footnote{e-mail:nassurmaeva@gmail.com}, A. V. Kpadonou$^{(a,c)}$\footnote{e-mail: vkpadonou@gmail.com}, M. E. Rodrigues$^{(b)}$\footnote{e-mail: esialg@gmail.com},  
M. J. S. Houndjo$^{(a,d)}$\footnote{e-mail:
sthoundjo@yahoo.fr} and J. Tossa$^{(a)}$\footnote{e-mail: joel.tossa@imsp-uac.org}

$^a$ \,{\it Institut de Math\'{e}matiques et de Sciences Physiques (IMSP)}\\
 {\it 01 BP 613,  Porto-Novo, B\'{e}nin}\\
$^{b}$\,{Faculdade de Ci\^encias Exatas e Tecnologia, Universidade Federal do Par\'a - Campus Universit\'ario de
Abaetetuba, CEP 68440-000, Abaetetuba, Par\'a, Brazil}\\

$^{c}$\, {\it Ecole Normale Sup\'erieure de Natitingou - Universit\'e de Parakou - B\'enin}\\ 

$^{d}$\,{\it Facult\'e des Sciences et Techniques de Natitingou - Universit\'e de Parakou - B\'enin} \\

\date{}

\end{center}
\begin{abstract}
We consider an exponential model within the so-called $f(T)$ theory of gravity, where $T$ denotes the torsion scalar. We focus our work on a cosmological feature of such a model, checking whether it may describe the de Sitter stage of the current universe through the analysis of the redshift $z$. Our results shows  that the model reproduces the de Sitter stage only for low-redshifts, where the perturbation function goes toward zero as the low values of the redshift are reached, whereas the effective parameter of equation of state goes to $-1$, which is the expected behavior for any model able to reproduce de Sitter stage.
\end{abstract}

Pacs numbers: 98.80.-k, 04.50.Kd, 95.36.+x

\section{Introduction}
According to the recent astronomical data from Type Ia supernovae \cite{1de1207.1646} as well as from the CMB spectrum \cite{2de1207.1646}, it is well known that our universe is experiencing an accelerating expanded phase. In order to comply with this feature, dark energy content for the universe is assumed \cite{3de1207.1646} within the framework of the General Relativity (GR), having undesired properties such as the violation of some energy conditions. Nowadays, it is a well-known fact that modifying the law of gravity leads to possible explanations  for the acceleration mechanism of the universe \cite{1dediego,2dediegof}. Various modified theory are proposed, one assumed as an algebraic function of the curvature $R$, the so-called $f(R)$ theory of gravity \cite{8a14deines1,dines}, other as an algebraic function for both the curvature and the Gauss-Bonnet invariant $G$ \cite{15a19deines1,dines2}.The recent and interesting kind of modified theory of gravity is the one base on the torsion scalar, as algebraic function of this later, denoted by $f(T)$ \cite{20a57deines1,dines3}.\par
The modified theory of gravity undertaken in this paper, called $f(T)$, is a modified version of the teleparallel (TT) essential based on the torsion scalar $T$. In this way, the torsion, as the TT term, is replaced by an algebraic function of the torsion, and this, always in the optic to comply with the cosmological data. In this paper, we are interested in the cosmological behavior of realistic models of modified gravity describing the de Sitter epoch of the current universe. As a check, we will study  homogeneous perturbations around the de Sitter solution of the dark energy density, to see whether we are able to regain the well known results. Note that the analysis is possible only the explicit form of the algebraic function is known. We therefore consider an exponential form of $f(T)$. Remember that this kind of model is well-known is the framework of $f(R)$  \cite{12de1108.6184v2}, where the cosmological evolution has been explored. More precisely in the framework of $f(T)$ theory of gravity, Bamba and collaborators considered an exponential model and studied cosmological evolutions of the equation of state for dark energy. They result shows that the crossing of the phantom divide line of $\omega_{DE}=-1$ may be observed when a logarithmic correction is performed to the exponential model. 
 In the paper \cite{perturbation3}, it is shown that for $\omega$ is pure-imaginary, the model is unstable. Also, it is shown that the model defined by the equation (39) is stable if $)<q<1$. On the other hand, in a  work developed by Bamba and collaborators, it has been shown that the model (39) of our paper should not produce ``finite time singularities"  \cite{1205.3421}. The well known of finite time singularities is the Big-Rip. Note that the Big-Rip appears only in the phantom phase. Therefore, the model under consideration here may just reproduce a quintessence universe, meaning that the greater value of the effective parameter of equation of state is $-1$. \par
In this paper, we still obey these features of the considered model, considering the suitable values of the parameter $q$ as pointed out in \cite{perturbation3}, and try to check whether such a model should describe the de Sitter stage of our current universe. This paper can be view as an extension of these cited paper, still in the optic to point out the how much this the exponential $f(T)$ model can be assumed as a viable one. To do so, we perform a perturbation around the de Sitter point and such for the behaviour of the perturbation function $y(z)$. We see that as the low red-shifts are being reached, the perturbation parameter goes to zero. This means that the exponential $f(T)$ can describe the de Sitter stage only for low red-shift. It is why for our study in this paper, we limit the values of $z$ to the lowest ones. It is can be view from our results that the behaviour of the parameter  of the equation of state related to the dark energy is the same as in the Bamba's paper \cite{1205.3421}, despite of using different values of the parameter $q$ belonging to the interval imposed by \cite{perturbation3}.
 Also, in this paper, within suitable values of the input parameters
 we plot the cosmological  parameters $\omega_{DE}$ and $\omega_{eff}$ and also the behaviour  of the perturbation function $y(z)$. Our results perfectly fit with the cosmological observational data and more precisely, for the low-redshift, the function $y(z)$ goes towards zero whereas $\omega_{eff}$ tends to $-1$, showing the stability of the model under consideration around the de Sitter solution.\par
The paper is organized as follows:  In $\ref{sec2}$ we present the generality on $f(T)$, putting up the general equation of motion and its associated generalized Friedmann equations from which all the analysis are realized.  The $\ref{sec3}$ is devoted for obtaining the general scale factor according to the model. The conclusion and perspective are presented in the last section.

\section{Generality on $f(T)$ theory}\label{sec2}
The modified theory of gravity based on the torsion scalar is the one for which the geometric part of the action is an algebraic function depending on the torsion. In the same way as in the Teleparallel gravity, the geometric elements are described using orthonormal tetrads components defined in the tangent space at each point of the manifold. In general the line element can be written as
\begin{eqnarray}
ds^2=g_{\mu\nu}dx^\mu dx^\nu=\eta_{ij}\theta^i\theta^j\,,
\end{eqnarray}
where we define the following elements
\begin{eqnarray}
dx^\mu=e_{i}^{\;\;\mu}\theta^{i}\,\quad \theta^{i}=e^{i}_{\;\;\mu}dx^{\mu}.
\end{eqnarray}
Note that $\eta_{ij}=diag(1,-1,-1,-1)$ is the metric related to the  Minkowskian spacetime and the $\{e^{i}_{\;\mu}\}$ are the components  of the tetrad which satisfy the following identity
\begin{eqnarray}
e^{\;\;\mu}_{i}e^{i}_{\;\;\nu}=\delta^{\mu}_{\nu},\quad e^{\;\;i}_{\mu}e^{\mu}_{\;\;j}=\delta^{i}_{j}.
\end{eqnarray}
The connection in use in this theory is the one of  Weizenbock's,  defined by
\begin{eqnarray}
\Gamma^{\lambda}_{\mu\nu}=e^{\;\;\lambda}_{i}\partial_{\mu}e^{i}_{\;\;\nu}=-e^{i}_{\;\;\mu}\partial_\nu e_{i}^{\;\;\lambda}.
\end{eqnarray}
Once the previous connection is assumed, one can then expression the main geometric objects; the torsion tensor's components as\begin{eqnarray}
T^{\lambda}_{\;\;\;\mu\nu}= \Gamma^{\lambda}_{\mu\nu}-\Gamma^{\lambda}_{\nu\mu},
\end{eqnarray}
which is used in the definition of the contorsion tensor as
\begin{eqnarray}
K^{\mu\nu}_{\;\;\;\;\lambda}=-\frac{1}{2}\left(T^{\mu\nu}_{\;\;\;\lambda}-T^{\nu\mu}_{\;\;\;\;\lambda}+T^{\;\;\;\nu\mu}_{\lambda}\right)\,\,.
\end{eqnarray}
The above objects (torsion and contorsion) are used to define a new tensor $S_{\lambda}^{\;\;\mu\nu}$ as
\begin{eqnarray}
S_{\lambda}^{\;\;\mu\nu}=\frac{1}{2}\left(K^{\mu\nu}_{\;\;\;\;\lambda}+\delta^{\mu}_{\lambda}T^{\alpha\nu}_{\;\;\;\;\alpha}-\delta^{\nu}_{\lambda}T^{\alpha\mu}_{\;\;\;\;\alpha}\right)\,\,.
\end{eqnarray}
The torsion scalar is defined from the previous tensor and the torsion tensor as
\begin{eqnarray}
T=T^{\lambda}_{\;\;\;\mu\nu}S^{\;\;\;\mu\nu}_{\lambda}
\end{eqnarray}

Let us write the action for the modified $f(T)$ theory with matter as follows
\begin{equation}
\label{1}
 S=\int d^{4}x e\left[\frac{f(T)}{2\kappa^{2}}+\mathcal{L}_{(matter)}\right],
\end{equation}
where $e\equiv\det[e^{i}\,_{\mu}]=\sqrt{-g}$ denotes the determinant of the tetrad, and $g$ the determinant of the space-time metric and the algebraic function $f(T)$ is assumed as the sum of the Teleparallel gravity term ( the torsion scalar $T$), and an additive algebraic function $F(T)$ as
\begin{equation}
\label{2}
 f(T)=T+F(T).
\end{equation}
But for the instance we will continue working in the general scheme using the algebraic function $f(T)$, having in view that the additive function $F(T)$  has to be used later.\par
By varying Eq.~$\ref{1}$ with respect to the tetrad, one gets the following general equation of motion \cite{ddbarrow}, 

\begin{eqnarray}
S^{\;\;\; \nu \rho}_{\mu} \partial_{\rho} T f_{TT} + 
[e^{-1} e^{i}_{\;\; \mu}\partial_{\rho}(e e^{\;\; \mu}_{i}S^{\;\;\; \nu\lambda}_{\alpha} )
+T^{\alpha}_{\;\;\; \lambda \mu}   S^{\;\;\; \nu \lambda}_{\alpha} ]f_{T}+
\frac{1}{4}\delta^{\nu}_{\mu}f=\frac{\kappa^{2}}{2} \mathcal{T}^{\nu}_{\mu},  \label{eq10}
\end{eqnarray}

where  $\mathcal{T}^{\nu}_{\mu}$ denotes  the stress tensor such that
\begin{eqnarray}
\mathcal{T}^{\nu}_{\mu}=\left(\rho+p\right)u_{\mu}u^{\nu}-p\delta^{\nu}_{\mu},
\end{eqnarray}
$\rho$ and $p$ being the energy density and the pressure of ordinary content of the universe respectively and $u^{\mu}$, the four-velocity such that $u^{\mu}u_{\mu}=1$. Still with the  considered signature (+,-,-,-), we assume the line element for the flat Robertson-Walker universe as
\begin{equation}
\label{4}
 dS^{2}=dt^{2}-a^{2}(t)d{\bf x}^{2},
\end{equation}
where $a(t)$ is the universe scalar factor. Therefore, the torsion scalar is performed giving 
\begin{equation}
\label{5}
 T=-6H^{2}.
\end{equation}

From Eq.$(\ref{eq10})$,   and using (\ref{4}),  one gets the following  gravitational field equations:
\begin{equation}
 \label{6}
 -Tf_{T}(T)+\frac{1}{2}f(T)=\kappa^{2}\rho,
 \end{equation}
 \begin{equation}
 \label{7}
 2{\dot T}Hf_{TT}(T)+2({\dot H}+3H^{2})f_{T}(T)+\frac{1}{2}f(T)=-\kappa^{2}p.
\end{equation}
Here the ``dot" denotes the derivative with respect to the cosmic time $t$, $H={\dot a}(t)/a(t)$, the  Hubble parameter  and
$\rho$ and $p$ , the matter energy-density and pressure, respectively.  The matter conservation equation reads 
\begin{equation}
\label{8}
 \dot\rho+3H(\rho+p)=0.
\end{equation}
By assuming that the matter content is a perfect fluid, the pressure and the energy density are related by the barotropic equation  of state (EoS)
\begin{equation}
\label{9}
 p=\omega\rho, 
\end{equation}
where $\omega$ is the  EoS-parameter for matter. For standard matter (non-relativist matter), $\omega_{m}=0$ and $\rho_{m}=\rho^{(0)}_{m}a(t)^{-3}$, whereas
for radiation (relativist matter), $\omega_{r}=1/3$ and $\rho_{r}=\rho^{(0)}_{r}a(t)^{-4}$.

Now, making use of the relation $f(T)=T+F(T)$, one can  write the Eqs.$(\ref{6})-(\ref{7})$ as functions of the effective energy density, $\rho_{eff}$, and effective pressure,
$p_{eff}$ as 
\begin{eqnarray}
\label{10}
 \rho_{eff}& = &\frac{3}{\kappa^{2}}H^{2},\\
 \label{11}
 p_{eff}& = &-\frac{1}{\kappa^{2}}\left( 2{\dot H}+3H^{2} \right),
\end{eqnarray}
where 
\begin{eqnarray}
 \label{12}
 \rho_{eff}& = &\rho-\frac{1}{2\kappa^{2}}\left[F(T)-2TF_{T}(T)\right],\\
 \label{13}
 p_{eff}& = &p+\frac{1}{\kappa^{2}}\left[F(T)+4\left({\dot H}+3H^{2}\right)F_{T}(T)+4H{\dot T}F_{TT}(T)\right].
\end{eqnarray}
Let us  define the dark energy density $\rho_{DE}$ as $\rho_{DE}=\rho_{eff}-\rho$ and introduce the variable 
\begin{equation}
\label{14}
y_{H}(z)\equiv\frac{\rho_{DE}}{\rho^{0}_{m}}=\frac{H^{2}}{\bar{m}^{2}}-(z+1)^{3}-\chi(z+1)^{4},
\end{equation}
where, $\rho^{(0)}_{m}$ is the energy density of matter at present time, $\bar{m}^{2}$ being the mass scale
\begin{equation}
\label{15}
 \bar{m}^{2}\equiv\frac{\kappa^{2}\rho^{0}_{m}}{3}\simeq1.5\times10^{-67}eV^{2},
\end{equation}
and $\chi$ defined by 
\begin{equation}
\label{16}
 \chi\equiv\frac{\rho^{0}_{r}}{\rho^{0}_{m}},
\end{equation}
where $\rho^{(0)}_{r}$ is the current radiation density, $z$ the redshift parameter, $z=1/a(t)-1$, and $y_{H}(z)$, written as  a function of $z$.
The EoS parameter for the dark energy, $\omega_{DE}$, is written as
\begin{equation}
\label{17}
 \omega_{DE}=-1+\frac{1}{3}(z+1)\frac{1}{y_{H}(z)}\frac{dy_{H}(z)}{dz}.
\end{equation}
By combining Eq.$(\ref{11})$ with Eq.$(\ref{5})$ and using Eq.$(\ref{14})$, one gets
\begin{equation}
\label{18}
 \frac{dy_{H}(z)}{dz}+J_{1}y_{H}(z)+J_{2}=0,
\end{equation}
where
\begin{equation}
\label{19}
 J_{1}=-\frac{3}{(z+1)}\frac{(1+2F_{T}(T))+4\bar{m}^{2}\left(3(z+1)^{3}+4\chi(z+1)^{4}\right)F_{TT}(T)}{(1+F_{T}(T))-12\bar{m}^{2}\left(y_{H}(z)+(z+1)^{3}+\chi(z+1)^{4}\right)F_{TT}(T)},
\end{equation}
\begin{eqnarray}
\label{20}
 J_{2}& = &\frac{1}{(z+1)\left(1+F_{T}(T)-12\bar{m}^{2}\left(y_{H}(z)+(z+1)^{3}+\chi(z+1)^{4}\right)F_{TT}(T)\right)}\times\nonumber\\
 &&\left\{\left(3(z+1)^{3}+4\chi(z+1)^{4}\right)\left((1+F_{T}(T))-12\bar{m}^{2}\left((z+1)^{3}+\chi(z+1)^{4}\right)F_{TT}(T)\right)\right.\nonumber\\
 &&\left.-3(1+2F_{T}(T))\left((z+1)^{3}+\chi(z+1)^{4}\right)-\left((F(T)/2\bar{m}^{2})+\chi(z+1)^{4}\right)\right\}.
\end{eqnarray}
The torsion scalar can be written in function of the red-shift as
\begin{equation}
\label{22}
T=-6\bar{m}^{2}\left(y_{H}(z)+(z+1)^{3}+\chi(z+1)^{4}\right),
\end{equation}
where we used $d/dt=-(z+1)H(z)d/dz=H(t)d/d(\ln a(t))$.  By taking the trace form of Eq.~$\ref{eq10}$ in the vacuum, one gets the following equation 
\begin{equation}
\label{23}
 3{\dot H}\left(2Tf_{TT}(T)+f_{T}(T)\right)-2Tf_{T}(T)+f(T)=0.
\end{equation}
Since we are considering the de Sitter universe, the torsion has to be taken as the de Sitter one, known as a constant, because of the constance of the Hubble parameter, such that the trace equation becomes
\begin{equation}
\label{24}
 f(T_{dS})-2T_{dS}f_{T}(T_{dS})=0,
\end{equation}
where $T_{dS}=constant$ is the de Sitter torsion scalar.\par
Now, in order to check how much  our model deviates from the de Sitter one, say, checking it convergence to the de Sitter model, we  study perturbations around the de Sitter solution of the dark energy density. Then, we write the following equation
\begin{equation}
\label{25}
 y_{H}(z)\simeq y_{0}+y(z),
\end{equation}
where $y_{0}=-T_{dS}/6\bar{m}^{2}$   is a constant and the stability  requires  $|y(z)|\ll1$.  Eq.$(\ref{22})$ leads to
\begin{equation}
\label{26}
 T=-6\bar{m}^{2}\left(y_{0}+y(z)+(z+1)^{3}+\chi(z+1)^{4}\right).
\end{equation}
In this case, by neglecting the contribution of radiation and taking into account only the non-relativist matter  contribution, but assumed to be much smaller 
than $y_{0}$, Eq.$(\ref{18})$ becomes
\begin{equation}
\label{27}
 \frac{dy}{dz}+\frac{\sigma}{(z+1)}y=\beta(z+1)^{2},
\end{equation}
where
\begin{equation}
\label{28}
 \sigma=-\frac{3\left(2f_{T}(T_{dS})-1\right)}{f_{T}+2T_{dS}f_{TT}(T_{dS})},
\end{equation}
\begin{equation}
\label{29}
 \beta=\frac{3\left(f_{T}(T_{dS})-1\right)}{f_{T}+2T_{dS}f_{TT}(T_{dS})}
\end{equation}

Here we have performed the variation of Eq.$(\ref{26})$ with respect to $T$, and also have used Eq.$(\ref{24})$.
The solution of Eq.$(\ref{27})$ is
\begin{equation}
 \label{31}
 y(z)=C_{0}(z+1)^{-\sigma}+\frac{\beta}{(3+\sigma)}(z+1)^{3},
\end{equation}
where $C_{0}$ is an integration  constant.\par
\begin{figure}
\begin{center}
\includegraphics[angle=0, width=0.5\textwidth]{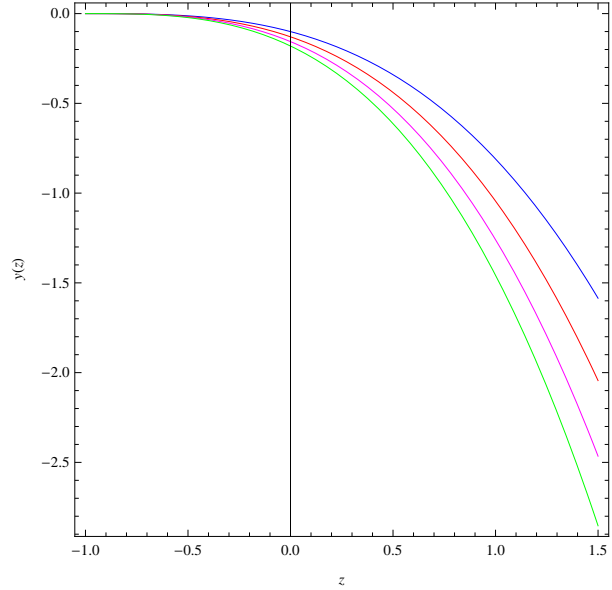}
\end{center}
\caption{\label{fig1} The curves plotted present the evolution of the perturbation function $y$  versus $z$. The Blue one is plotted  for $q=0.03$, the Red for $q=0.04$, the Magenta for $q=0.05$ and the Green one for $q=0.06$. The parameters $C_0$ and $\Omega_{m}^{(0)}$ are set to $0.5$ and $0.26$, respectively.}
\end{figure}
\begin{figure}
\begin{center}
\includegraphics[angle=0, width=0.5\textwidth]{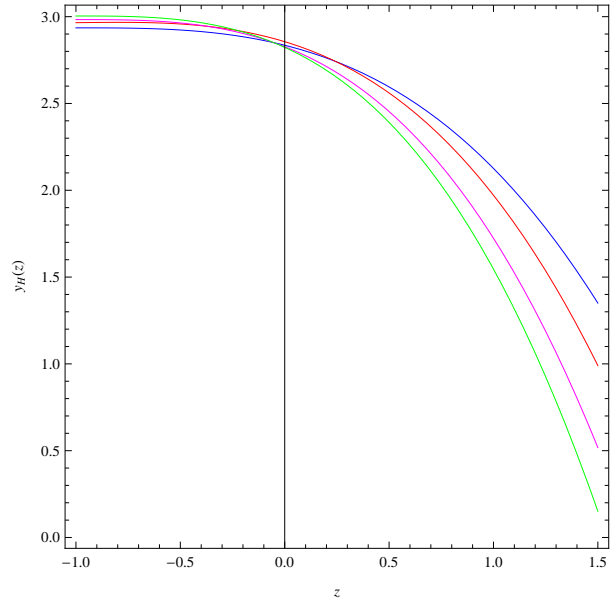}
\end{center}
\caption{\label{fig2} The graph presents the evolution of $y_H$ versus $z$ for a wide interval of the redshift. The Blue one is plotted for $q=0.03$, the Red for $q=0.04$, the Magenta for $q=0.05$ and the Green one for $q=0.06$. All of them are plotted for $C_0=0.5$,  $\Omega_{m}^{(0)}=0.26$. }
\end{figure}

As a check for our approach, let us now consider an exponential model, to see the behaviour of the dark energy. In this paper, we assume the following exponential model;  $f(T)=T+\alpha T\left(1-e^{qT_0/T}\right)$. Note that this model has been early used by Bamba and collaborators in \cite{bamba1} where they studied the  cosmological evolutions of the equation of state for dark energy. They showed that the crossing of the phantom divide line of $\omega_{DE}=-1$ can be realized by combining to the exponential model a logarithmic one. In this section, our goal is just to check the convergence of the exponential model to the de Sitter one. In this way, we have

\begin{equation}
\label{48}
 F(T)=\alpha T\left(1-e^{qT_0/T}\right),
\end{equation}
where $T_{0}=-6H_0^2$ is the scalar tensor at present time, with $H_0=74.2\pm 3.6 Km s^{-1} Mpc^{-1}$, and $\alpha=-(1-\Omega_m^{(0)})/\left(1-\left(1-2q\right)e^q\right)$. By considering $ T_{dS}/T_{0}\ll1$, and using Eq~(\ref{24}) within  the expansion of the exponential function, one gets the following expression for the de Sitter torsion
\begin{eqnarray}
T_{dS}=-\frac{qT_0}{2}\left(\alpha-\sqrt{\alpha^2 -8\alpha}\right)
\end{eqnarray} 
In order to reach to an interesting physical result, among other conditions, we assume the case where $q>0$ and $\alpha<0$. Therefore, one gets 
\begin{eqnarray}
\sigma=\frac{-3\left[2+\frac{1}{2}(\alpha-\sqrt{\alpha^{2}-8\alpha}-4)\right]}
 {\frac{1}{2}(\alpha-\sqrt{\alpha^{2}-8\alpha}-4)+\frac{4}{\alpha-\sqrt{\alpha^{2}-8\alpha}}-1}, \\
 \beta=\frac{3}{\frac{1}{2}(\alpha-\sqrt{\alpha^{2}-8\alpha}-4)+\frac{4}{\alpha-\sqrt{\alpha^{2}-8\alpha}}-1},
\end{eqnarray}
and the perturbation function $y(z)$ takes the following form
\begin{eqnarray}
 y(z)=c_{0}(z+1)^{-\sigma}
 -\frac{1}{3-\frac{4}{\alpha-\sqrt{\alpha^{2}-8\alpha}}}(z+1)^{3},
\end{eqnarray}
and whose the evolution versus redshift is realized at the Fig. {\bf \ref{fig2}}
From the equation (\ref{25}), we obtain the corresponding expression of $y_H(z)$ as
\begin{eqnarray}
 y_{H}(z)=\frac{pT_{0}}{12\bar{m}^{2}}\left(\alpha-\sqrt{\alpha^{2}-8\alpha}\right)+c_{0}(z+1)^{-\sigma}
 -\frac{1}{3-\frac{4}{\alpha-\sqrt{\alpha^{2}-8\alpha}}}(z+1)^{3}.
\end{eqnarray}
The graphs representing the evolution of this function are plotted at the Fig. {\bf \ref{fig1}}. Here, we just try to show what happens at the low-redshift level. It can be observed that as the lowest redshift are being reached, i.e, $z\rightarrow -1^{+}$, the curves at the right hand side seem to go toward zero. This is just the effect for having used a very large interval, i.e, $]-1,+\infty[$ because, indeed, at the level of low-redshift the curved tend to some values different from zero. \par
\begin{figure}
\begin{center}
\includegraphics[angle=0, width=0.5\textwidth]{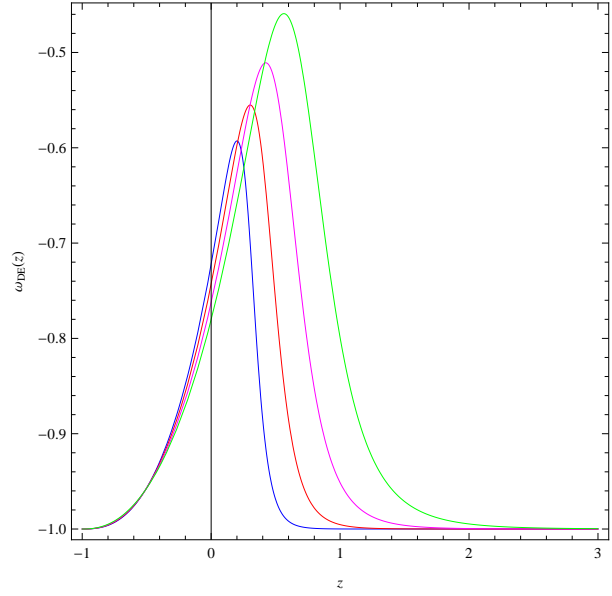}
\end{center}
\caption{\label{fig3} The graph presents the evolutions of the parameter of the equation of state related to the dark energy according to the physical evolution of the redshift. The Blue one is plotted for $q=0.03$, the Red for $q=0.04$, the Magenta for $q=0.05$ and the Green one for $q=0.06$. The parameters $C_0$ and $\Omega_{m}^{(0)}$ are set to $0.5$ and $0.26$, respectively.}
\end{figure}

\begin{figure}
\begin{center}
\includegraphics[angle=0, width=0.5\textwidth]{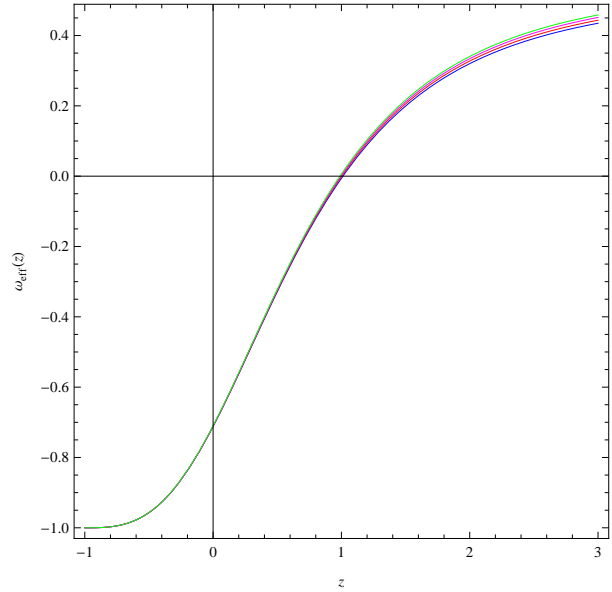}
\end{center}
\caption{\label{fig4} This graph presents the evolutions of the parameter of the effective equation of state versus $z$, the blue for $q=0.03$, the Red for $q=0.04$, the Magenta for $q=0.05$ and the Green one for $q=0.06$. Also here $C_0$ and $\Omega_{m}^{(0)}$ are set to $0.5$ and $0.26$, respectively.}
\end{figure}

 
Within the above expression, we perform the ones of the parameters of equation of state of the dark energy $\omega_{DE}$  and the effective $\omega_{eff}$ as
\begin{eqnarray} 
\omega_{DE}&=&-1-\frac{\sigma\left(3+\sigma\right)\left(z+1\right)^{-\sigma}-3\beta\left(z+1\right)^3}{3\left(3+\sigma\right)y_H(z)},\\
\omega_{eff}&=&-1-\frac{\sigma\left(3+\sigma\right)\left(z+1\right)^{-\sigma}-3\left(3+\sigma+\beta\right)\left(z+1\right)^3}{3\left(3+\sigma\right)\left[y_H(z)+\left(z+1\right)^3\right]},
\end{eqnarray}
whose the graphs representatives are plotted respectively at Fig. {\bf \ref{fig3}} and Fig. {\bf \ref{fig4}} .
We see from the Fig. {\bf \ref{fig3}} that as the lowest redshift are being reached, the parameter $\omega_{DE}$ tends to $-1$ for the values of the input parameters $q$ and $\alpha$.  The same is almost obtained for the effective parameter of EoS $\omega_{eff}$ showing for any of the considered values of the input parameters $q$, the curves go toward $-1$. This is an interesting result. Note that the effective parameter $\omega_{eff}$ is the one that characterizes the whole content of the universe. Also, it is well known that the de Sitter universe is the one for which one has $\omega_{eff}=-1$. Therefore, due to fact that  $\omega_{eff}\rightarrow -1$  for $z\rightarrow -1^{+}$, one can conclude that within the exponential $f(T)$ model, the de Sitter stage of the universe is realized for the lowest values of the redshift. This result completely agrees with the one obtained by Bamba and collaborators \cite{bamba1} where they showed within others values of the input parameters that the parameter $\omega_{DE}$ goes toward $-1$ as the lowest redshift are being reached $z\rightarrow -1^{+}$ with an exponential $f(T)$ model. In our case, due to the fact that we approximate the dark energy to the effective content of the universe, it is obvious that our result is in agreement with the one of \cite{bamba1}. \par
 We point out that our study is just about low red-shift. Note that as the title mentions, we are interested to see whether the exponential model should describe the de Sitter stage of our current universe. Therefore, we just have to limit our analysis to the low. This king of study is in the same way as the well know $\Lambda CDM$ model, which works only for the current times. It is well known that for early times, $\Lambda CDM$ model does not work. The spirit is just the same here. We just  our study to the low red-shift because knowing that this cannot lead any interesting result for high red-shift.

%%%%%%%%%%%%%%%%%%%%%%%%%%%%%%%%%%%%%%%%%%%%%%%%%%%%%%%
%%%%%%%%%%%%%%%%%%%%%%%%%%%%%%%%%%%%%%%%%%%%%%%%%%%%%%
\section{Time evolution}\label{sec3}
%%%%%%%%%%%%%%%%%%%%%%%%%%%%%%%%%%%%%%%%%%%%%%%%%%%%%%%%%%%%%%%%%%%%%%%%%%%%%%%%%%%%%%%%%%%%%%%
Our goal in this section in to obtain the expression of the scale factor for with, at the level of low-redshift, the standard de Sitter expression may be recovered. To do this, let us consider the Eqs.$(\ref{25})$ and $(\ref{31})$ in the case where $-\sigma=\gamma$ is a positive real number:
\begin{equation}
 \label{42}
 y_{H}(z)=y_{0}+C_{0}(z+1)^{\gamma}+\frac{\beta}{(3+\alpha)}(z+1)^{3},
\end{equation}
where $y_{0}=H^{2}_{dS}/{\bar{m}}^{2}=-T_{dS}/6{\bar{m}}^{2}$ and $C_{0}$ is a constant. We will assume $C_{0}>0$.
From the first equation of motion $(\ref{10})$, one gets
\begin{equation}
 \label{43}
 \frac{H^{2}}{{\bar{m}}^{2}}=y_{H}(z)+(z+1)^{2}\equiv y_{0}+C_{0}(z+1)^{\gamma}+\frac{1}{1+\frac{1}{2T_{dS}f_{TT}(T_{dS})}(1-f_{T}(T_{dS}))}(z+1)^{3},
\end{equation}
and the explicit expression of $H$ depending on cosmic time $t$ may be obtained. By substituting  $(z+1)$ by $1/a(t)$, one gets
\begin{equation}
 \label{44}
 \left(\frac{{\dot a}(t)}{a(t)}\right)=H^{2}_{dS}+(C_{0}{\bar{m}}^{2})\left(\frac{1}{a(t)}\right)^{\gamma}.
\end{equation}
In the previous expression we have neglected the matter contribution. Now by considering $t>0$, the general solution for the expanding universe is
\begin{equation}
 \label{45}
 a(t)=\left(\frac{C_{0}{\bar{m}}^{2}}{H^{2}_{dS}}\right)^{\frac{1}{\gamma}}\left[\sinh\left(\frac{H_{dS}}{2}\gamma t+\phi\right)\right]^{\frac{2}{\gamma}},
\end{equation}
where $\phi$ is a positive constant. The above expression can be rewritten, given
\begin{equation}
 \label{46}
 a(t)=a_{0}e^{H_{dS}t}\left[1-e^{-(H_{dS}\gamma t+2\phi)}\right]^{\frac{2}{\gamma}},
\end{equation}
where $a_{0}$ is the value of the scalar factor $a(t)$ at $t=0$ and $a_{0}=\frac{1}{2}\left(\frac{C_{0}\bar{m}^{2}e^{2\phi}}{H_{dS}}\right)^{\frac{1}{\gamma}}$. Therefore, the Hubble parameter gives
\begin{equation}
 \label{47}
 H=H_{dS}\coth\left(\frac{H_{dS}}{2}\gamma t+\phi\right).
\end{equation}
The interesting feature here is to observe that, as $\gamma>0$, for the the redshifts reaching the low possible ones, i.e, $z\rightarrow -1^{+}$,  the scale factor becomes 
$a(t)\simeq a_{0}e^{H_{dS}t}$, and  
for $t>0$, $H\simeq H_{dS}$. These results, once again confirm that the de Sitter stage of the current dark energy dominated universe may be described for the lowest values for the redshift. 
\vspace{4cm}

%\begin{center}
%\begin{figure}[htbp]
%\begin{minipage}[t]{1.00\linewidth}
%\includegraphics[width=\linewidth]{y[z].eps}
%\end{minipage} \hfill
%\caption{{\protect\footnotesize 
%The curves plotted present the evolution of the perturbation function $y$  versus $z$. The Blue one is plotted  for $q=0.03$, the Red for $q=0.04$, the Magenta for $q=0.05$ and the Green one for $q=0.06$. The parameters $C_0$ and $\Omega_{m}^{(0)}$ are set to $0.5$ and $0.26$, respectively.  }}
%\label{fig1}
%\end{figure}
%\end{center}

%\begin{figure}[htbp]
%\centering
%\begin{tabular}{rl}
%\includegraphics[height=7cm,width=7cm]{yH1[z].eps} \label{e1}&
%\includegraphics[height=7cm,width=7cm]{yH2[z].eps} \label{e2}
%\end{tabular}
%\caption{{\protect\footnotesize }}
%\label{fig1}
%\end{figure}

%\begin{center}
%\begin{figure}[htbp]
%\begin{minipage}[t]{1.00\linewidth}
%\includegraphics[width=\linewidth]{yH.eps}
%\end{minipage} \hfill
%\caption{{\protect\footnotesize 
%The two graphs present both the evolution of $y_H$ versus $z$ for different intervals of the redshift. The Blue one is plotted for $q=0.03$, the Red for $q=0.04$, the Magenta for $q=0.05$ and the Green one for $q=0.06$. We set the paprameters  $C_0=0.5$ and  $\Omega_{m}^{(0)}=0.26$.   We mention here that the right hand side one is plotted for $z \in ]-1,+\infty[$.  }}
%\label{fig2}
%\end{figure}
%\end{center}

%%%%%%%%%%%%%%%%%%%%%%%%%%%%%%%%%%%%%%%%%%%%%%%%%%%%%%%%%%
%%%%%%%%%%%%%%%%%%%%%%%%%%%%%%%%%%%%%%%%%%%%%%%%%%%%%%%%%%

%\begin{figure}[htbp]
%\centering
%\begin{tabular}{rl}
%\includegraphics[height=7cm,width=7cm]{yH1[z].eps} \label{e1}&
%\includegraphics[height=7cm,width=7cm]{yH2[z].eps} \label{e2}
%\end{tabular}
%\caption{ }
%\label{fig2}
%\end{figure}

%\begin{center}
%\begin{figure}[htbp]
%\begin{minipage}[t]{1.00\linewidth}
%\includegraphics[width=\linewidth]{WDE[z].eps}
%\end{minipage} \hfill
%\caption{{\protect\footnotesize 
%The graph presents the evolutions of the parameter of the equation of state related to the dark energy according to the physical evolution of the redshift. The Blue one is plotted for $q=0.03$, the Red for $q=0.04$, the Magenta for $q=0.05$ and the Green one for $q=0.06$. The parameters $C_0$ and $\Omega_{m}^{(0)}$ are set to $0.5$ and $0.26$, respectively. }}
%\label{fig3}
%\end{figure}
%\end{center}

%\begin{center}
%\begin{figure}[htbp]
%\begin{minipage}[t]{1.00\linewidth}
%\includegraphics[width=\linewidth]{Weff.eps}
%\end{minipage} \hfill
%\caption{{\protect\footnotesize 
%This graph present the evolutions of the parameter of the effective equation of state versus $z$, the blue for $q=0.03$, the Red for $q=0.04$, the Magenta for $q=0.05$ and the Green one for $q=0.06$. Also here $C_0$ and $\Omega_{m}^{(0)}$ are set to $0.5$ and $0.26$, respectively.  }}
%\label{fig4}
%\end{figure}
%\end{center}

%%%%%%%%%%%%%%%%%%%%%%%%%%%%%%%%%%%%%%%%%%%%%%%%%%%%%%%%%%%%%%%%%%%%%%%%%%%%%%%%%%%%%%%%%%%%
%%%%%%%%%%%%%%%%%%%%%%%%%%%%%%%%%%%%%%%%%%%%%%%%%%%%%%%%%%%%%%%%%%%%%%%%%%%%%%%%%%%%%%%%%%%%%

\newpage

%%%%%%%%%%%%%%%%%%%%%%%%%%%%%%%%%%%%%%%%%%%%%%%%%%%%%%%%%%%%%%%%%%%%%%%%%%%%%%%%%%%%%%%%%%%%%%%%
\section{Conclusion}
%%%%%%%%%%%%%%%%%%%%%%%%%%%%%%%%%%%%%%%%%%%%%%%%%%%%%%%%%%%%%%%%%%%%%%%%%%%%%%%%%%%%%%%%%%%%%%%%%%
In this paper, we consider the modified theory of gravity based on the torsion scalar, as the generalized version of the TT, where we assume an exponential model for the action. We then search whether the model may be used to describe the de Sitter stage of the current dark energy. To do so, the function $y_H(z)$, as the ratio of the energy density of the dark energy over the current ordinary matter, and perturb it around its de Sitter value. What has to be checked is whether within the exponential model, the perturbation $y(z)$ function may vanish or converge toward a value whose the magnitude should be very less than 1.\par
According to our results, the perturbation function $y(z)$ converges and goes toward zero as the lowest values ($z=-1^{+}$) are approached. As it is well known, for the de Sitter universe, the parameter of the effective equation of motion is $-1$. Beside to the interesting convergence of the perturbation function, we see that for the same values of the redshift, the parameter of the effective equation of state $\omega_{eff}$ goes toward $-1$, which is its known value of the de Sitter universe. At the same time, the parameter $y_H(z)$ reaches the corresponding de Sitter values $y_0$, as explicitly presented at the left hand side of the Fig. {\bf \ref{fig1}}.  In order to check the validity of the fact to get stability only for low-redshifts, we first try to perform the time evolution expression for both the scale factor and the Hubble parameter. The interesting feature here is that for the lowest values of $z$ and within the cosmological considerations, we see that the de Sitter expressions are recovered for the scale factor and the Hubble factor, confirming once again that the model can describe the de Sitter stage of the current accelerating expanded universe only for lowest values of $z$.
From ours results, we conclude that the exponential $f(T)$ model, for suitable values of the input parameters may perfectly describe the de Sitter stage of the current dark energy dominated universe.

\vspace{1.5cm}

{\bf Acknowledgement}: B. S. B. Nassur Maeva thanks a lot  DAAD for financial support.


\begin{thebibliography}{17}
\addcontentsline{toc}{chapter}{Bibliographie}

\bibitem{1de1207.1646} T. M. Davis et al., Astrophys. J. \textbf{666}, 716 (2007) [arXiv: astro-ph/0701510].

\bibitem{2de1207.1646} Dunkley et al [WMAP Collaboration], collaborations: arXiv: 0803.0586 [astro-ph], E. Komatsu et al. [WMAP Collaboration], arXiv: 0803.0547 [astro-ph].

\bibitem{3de1207.1646} T. Padmanabhan, AIP Conf. Proc. \textbf{861}, 179 (2006) arXiv: astro-ph/0603114.

\bibitem{1dediego} A. G. Riess et al. [Supernova Search Team Collaboration], Astron. J. \textbf{116}: 1009 (1998); 

\bibitem{2dediego} Edmund J. Compeland, M. Sami and Shinji Tsujikawa, Int. J. Mod. Phys. D. \textbf{15} (2006) 1753-1936 [arXiv: hep-th/0603057v3].

\bibitem{ddd} T. Padmanabhan, AIP Conf. Proc. \textbf{843}, 111 (2006) [astro-ph/0602117].

\bibitem{ddd} J. Frieman, M. Turner and D. Huterer, Ann. Rev. Astron. Astrophys. \textbf{46}, 385 (2008) [arXiv: 0803.0982[astro-ph]].

\bibitem{2dediegof} 
P. J. E. Peebles and B. Ratra, Rev. Mod. Phys. \textbf{75}, 559 (2003) [astro-ph/0207347].
%%%%%%%%%%%%%%%%%%%%%%%%%%%%%%%%%%%%%%%%%%%%%%%%%%%%%
%%%%%%%%%%%%%%%%%%%%%%%%%%%%%%%%%%%%%%%%%%%%%%%%%%%%
\bibitem{8a14deines1} S. Nojiri and S. D. Odintsov, ECONF C \textbf{0602061}, 06 (2006); Int. J. Geom. Meth. Mod. Phys. 4, 115-146 (2007) [arXiv: hep-th/0601213]; Phys. Rept. \textbf{505}, 59-144 (2011) [arXiv: 1011.0544].

\bibitem{dd} T. Harko, F. S. N. Lobo, S. Nojiri and S. D. Odintsov, Phys. Rev. D \textbf{84} (2011) 024020 [arXiv: 1104.2669[gr-qc]].

\bibitem{dd}
M. J. S. Houndjo, Int. J. Mod. Phys. D. \textbf{21}, 1250003 (2012). arXiv: 1107.3887 [astro-ph.CO].

\bibitem{dd} M. J. S. Houndjo and O. F. Piattella, Int. J. Mod. Phys. D. \textbf{21} 1250024 (2012) [arXiv: 1111.4275 [gr-qc]].

\bibitem{dd} D. Momeni, M. Jamil and R. Myrzakulov, Euro. Phys. J. C. \textbf{72}, [arXiv: 1107.5807 [Physics.gen-ph]].

\bibitem{dd}
M. J. S. Houndjo, C. E. M. Batista, J. P. Campos and O. F. Piattella,   Can. J. Phys. {\bf 91}, 548-553 (2013), [arXiv: 1203.6084[gr-qc]].



\bibitem{dines} F. G. Alvarenga, M. J. S. Houndjo, A. V. Monwanou and Jean B. Chabi-Orou, Journal of Modern Physics {\bf 4}, 130-139 (2013), [arXiv: 1205.4678[gr-qc]].

%%%%%%%%%%%%%%%%%%%%%%%%%%%%%%%%%%%%%%%%%%%%%%%%%%%%%
%%%%%%%%%%%%%%%%%%%%%%%%%%%%%%%%%%%%%%%%%%%%%%%%%%%%
\bibitem{15a19deines1} S. Nojiri and S. D. Odintsov, Phys. Lett. B. \textbf{631}, 1 (2005), [hep-th/0508049].

\bibitem{dd} S. Nojiri, S. D. Odintsov, A. Toporensky and P. Tretyakov, [arXiv: 0912.2488].

\bibitem{dd} K. Bamba, S. D. Odintsov, L. Sebastiani and S. Zerbini, [arXiv: 0911.4390].

\bibitem{dd} K. Bamba, C-Q. Geng, S. Nojiri and S. D. Odintsov, [arXiv: 0909.4397].

\bibitem{dd}
M. E. Rodrigues, M. J. S. Houndjo, D. Momeni and R. Myrzakulov, Can. J. Phys, {\bf 92}, 173-176 (2014), [arXiv: 1212.4488].

\bibitem{dines2} M. J. S. Houndjo, M. E. Rodrigues, D. Momeni and R. Myrzakulov, Accepted for publication in Can. J. Phys, [arXiv: 1301.4642[gr-qc]].

%%%%%%%%%%%%%%%%%%%%%%%%%%%%%%%%%%%%%%%%%%%%%%%%%%%%%
%%%%%%%%%%%%%%%%%%%%%%%%%%%%%%%%%%%%%%%%%%%%%%%%%%%%%
%%%%%%%%%%%%%%%%%%%%%%%%%%%%%%%%%%%%%%%%%%%%%%%%%%%%
\bibitem{20a57deines1} J. Amor{\'o}s, J. de Haro and S. D. Odintsov, Physical Review D {\bf 87}, 104037 (2013) [arXiv:1305.2344[gr-qc]].

\bibitem{dd} K. Bamba, J. de Haro and S. D. Odintsov,  JCAP \textbf{1302} (2013) 008 [arXiv:1211.2968[gr-qc]].


\bibitem{dd} K. Bamba, S. Nojiri and S. D. Odintsov, [arXiv: 1304.6191[gr-qc]].

\bibitem{dd} G. R. Bangochea, R. Ferraro, Phys. Rev. D. \textbf{79}, 124019 (2009) [arXiv: 0812.1205[gr-qc]].

\bibitem{dd} E. V. Linder, Phys. Rev. D \textbf{81} 127301 (2010) [Erratum-ibid D {\bf 82}, 109902 (2010)][arXiv: 1005.3039 [astro-ph.CO]].


 M. Jamil, D. Momeni and R. Myrzakulov, Eur.Phys. J. C. \textbf{72} (2012) 2267 [arXiv: 1212.6017[gr-qc]].
 
\bibitem{dd}  R. Myrzakulov, Entropy \textbf{14} (2012) 1627 [arXiv: 1212.2155[gr-qc]].

\bibitem{dd}
M. R. Setare and N. Mohammadipour, JCAP \textbf{1211} (2012) 030 [arXiv: 1211.1375[gr-qc]]; JCAP \textbf{01} (2013) 015 [arXiv: 1301.4891[gr-qc]].

\bibitem{dd} M. Jamil, D. Momeni, R. Myrzakulov and P. Rudra, J. Phys. Soc. Jap. \textbf{81} (2012) 114004 [arXiv: 1211.0018[Physics.gen-ph]].

\bibitem{dd} M. E. Rodrigues, M. J. S. Houndjo, D. Saez-Gomez and F. Rahaman, Phys. Rev. D. \textbf{86} (2012) 104059 [arXiv: 1209.4859[gr-qc]].

\bibitem{dd}
M. Jamil, D. Momeni and R. Myrzakulov, Eur. Phys. J. C \textbf{72} (2012) 2122 [arXiv: 1209.1298[gr-qr]].

\bibitem{dd} R. Myrzakulov, Eur.Phys. J. C \textbf{72} (2012) 2203 [arXiv: 1207.1039[gr-qc]].

\bibitem{dd} M. J. S Houndjo, D. Momeni and R. Myrzakulov, Int. J. Mod. Phys. D \textbf{21} (2012) 1250093 [arXiv: 1206.3938[Physics.gen-ph]].


\bibitem{dd} M. E. Rodrigues, M. H. Daouda and M. J. S. Houndjo, [arXiv: 1205.0565[gr-qc]].

\bibitem{dd} M. R. Setare and M. J. S. Houndjo, Can. J. Phys. {\bf 91} (2013) 260-267,[arXiv: 1203.1315[gr-qc]]. 



\bibitem{dd} S. Nesseris, S. Basilakos, E. N. Saridakis, L. Perivolaropoulos, Phys. Rev. D {\bf 88}, 103010 (2013); arXiv:1308.6142 [astro-ph.CO].

\bibitem{dd} A. Paliathanasis, S. Basilakos, E.N. Saridakis, U. Valparaiso, S. Capozziello, K. Atazadeh, F. Darabi and M. Tsamparlis, arXiv:1402.5935 [gr-qc].

\bibitem{dd}  S. Capozziello, P. A. Gonzalez, Emmanuel N. Saridakis and Yerko Vasquez, JHEP {\bf 1302}, (2013) 039; arXiv:1210.1098 [hep-th]. 

\bibitem{dd}    [pdf, ps, other]
Solar system constraints on f(T) gravity
Lorenzo Iorio, Emmanuel N. Saridakis, Mon. Not. Roy. Astron. Soc. {\bf 427} (2012) 1555; arXiv:1203.5781 [gr-qc].

\bibitem{dd}  Yi-Fu Cai, Shih-Hung Chen, J. B. Dent, S. Dutta, E. N. Saridakis, Quantum Grav. {\bf 28}, (2011) 215011; arXiv:1104.4349 [astro-ph.CO].

\bibitem{dd}  J. B. Dent, S. Dutta, E. N. Saridakis, JCAP {\bf 1101}, 009 (2011); arXiv:1010.2215 [astro-ph.CO].

\bibitem{dd} Shih-Hung Chen, J. B. Dent, S. Dutta, E. N. Saridakis, Phys. Rev. D {\bf 83} , 023508 (2011),  arXiv:1008.1250 [astro-ph.CO].


\bibitem{dd} K. Bamba, M. Jamil, D. Momeni and R. Myrzakulov, [arXiv: 1202.6114[Physics.gen-ph]].


\bibitem{dd} K. Bamba, S. 'i. Nojiri and S. D. Odintsov, Phys. Rev. D \textbf{85} (2012) 104036 [arXiv: 1202.4057[gr-qc]].

\bibitem{dd} M. Jamil, D. Momeni and R. Myrzakulov, Eur. Phys. J. C \textbf{72} (2012) 2267 [arXiv: 1212.6017[gr-qc]];  Gen. Rel. Grav. \textbf{45} (2013) 263 [arXiv: 1211.3740[physics.gen-ph]]; Eur. Phys. J. C \textbf{72} (2012) 2122 [arXiv: 1209.1298[gr-qc]];  Eur. Phys. J. C \textbf{72} (2012) 2075 [arXiv: 1208.0025[gr-qc]].

\bibitem{dd}  M. Jamil, K. Yesmakhanova, D. Momeni and R. Myrzakulov, Central Eur. J. Phys. \textbf{10} (2012) 1065 [arXiv: 1207.2735[gr-qc]].

\bibitem{dd}  M. J. S. Houndjo, D. Momeni and R. Myrzakulov, Int. J. Mod. Phys.  D \textbf{21} (2012) 1250093 [arXiv: 1206.3938[physics.gen-ph]].

\bibitem{dd}
M. Jamil, D. Momeni and R. Myrzakulov, Eur. Phys. J. C \textbf{72} (2012) 1959 [arXiv: 1202.4926[physics.gen-ph]].

\bibitem{dd} M. H. Daouda, M. E. Rodrigues and M. J. S. Houndjo, Phys. Lett. B \textbf{715} (2012) 241 [arXiv: 1202.1147[gr-qc]].

\bibitem{dd}  M. Jamil, S. Ali, D. Momeni and R. Myrzakulov, Eur. Phys. J. C \textbf{72} (2012) 1998 [arXiv: 1201.0895[physics.gen-ph]].

\bibitem{dd}  M. Jamil, D. Momeni N. S. Serikbayev and R. Myrzakulov, Astrophys. Space Sci. \textbf{339}, 37, (2012) [arXiv: 1112.4472[physics.gen-ph].

\bibitem{dd}
M. Jamil, D. Momeni and M. A. Rachid, Eur. Phys. J. C \textbf{71}, 1711 (2011) [arXiv: 1107.1558[physics.gen-ph]].

\bibitem{dd} M. Hamani Daouda, M. E. Rodrigues and M. J. S. Houndjo, Eur. Phys. J. C \textbf{72} (2012) 1893 [arXiv: 1111.6575[gr-qc]];  Eur. Phys. J. C \textbf{72} (2012) 1890 [arXiv: 1109.0528[physics.gen-ph]]; Eur. Phys. J. C \textbf{71} (2011) 1817 [arXiv: 1108.2920[astro-ph.CO]]. 

\bibitem{dd} R. Myrzakulov, Gen. Rev. Grav. \textbf{44} (2012) 3059 [arXiv: 1008.4486[physics.gen-ph]].

\bibitem{dd}
K. K. Yerzhanov, S. R. Myrzakul, I. I. Kulnazarov and R. Myrzakulov, [arXiv: 1006.3879[gr-qc]].


\bibitem{dd} R. Myrzakulov, Eur. Phys. J. C \textbf{71} (2011) 1752 [arXiv: 1006.1120[gr-qc]].


\bibitem{dd} M. E. Rodrigues, M. J. S. Houndjo, D. Momeni and R. Myrzakulov, [arXiv: 1302.43.72[physics.gen-ph]].

\bibitem{dd} J. M. Bardeen, B. Carter and S. W. Hawking, Math. Phys. \textbf{31} (1973) 161-170.

\bibitem{dd} 
N. Tamanini and C. G. Boehmer, Pyhs. Rev. D \textbf{86} 044009 (2012) [arXiv: 1204.4593[gr-qc]].

\bibitem{ddbarrow} Baojiu Li, T. P. Sotiriou and J. D. Barrow, Phys. Rev. D \textbf{83}, 064035 (2011); Phys. Rev. D \textbf{83}, 104030 (2011).

\bibitem{dd} 
M. J. S. Houndjo, D. Momeni, R. Myrzakulov and M. E. Rodrigues, [arXiv: 1304.1147 [physics.gen-ph]].


 \bibitem{dines3} C. Deliduman and B. Yapiskan, [arXiv: 1103.2225v2[gr-qc]].
 %%%%%%%%%%%%%%%%%%%%%%%%%%%%%%%%%%%%%%%%%%%%%%%%%%%
 %%%%%%%%%%%%%%%%%%%%%%%%%%%%%%%%%%%%%%%%%%%%%%%%%%%
\bibitem{12de1108.6184v2} K. Bamba, C. Q. Geng and C. C. Lee, J CAP \textbf{1008}, 021 (2010); [arXiv: 1005.4572[astro-ph.CO]].

\bibitem{bamba1} K. Bamba, C. Q. Geng, C. C. Lee and L. L. Luo, JCAP {\bf 1101}, 021  (2011); [arXiv: 1011. 0508v2[astro-ph.CO]].

\bibitem{1205.3421}  Kazuharu Bamba, Salvatore Capozziello, Shin'ichi Nojiri and Sergei D. Odintsov;[arXiv: 1205. 3421v3[gr-qc]]

\bibitem{perturbation3}  Shinh-Hung Chen, James B. Dent, Sourish Dutta, and Emmanuel N. Saridakis;[arXiv:1008. 1250v2[astro-ph.CO]]

\bibitem{bamba}   K. Bamba, C. Q. Geng, C. C. Lee and L. L. Luo, JCAP {\bf 1101}, 021  (2011); [arXiv: 1011. 0508v2[astro-ph.CO]]




\end{thebibliography}
\end{document}